\documentclass{iaus_298}
\usepackage{graphicx}

\title[C and O abundances in stellar populations] 
{The evolution of C and O abundances in stellar populations}

\author[P. E. Nissen \& W. J. Schuster]   
{Poul E. Nissen$^1$
 \and William J. Schuster$^2$}

\affiliation{$^1$ Stellar Astrophysics Centre, Dept. of Physics and Astronomy,
Aarhus University, Denmark\\ email: {\tt pen@phys.au.dk} \\[\affilskip]
$^2$Observatorio Astronomico National, UNAM, Ensenada, 
B.C. Mexico \\email: {\tt schuster@astrosen.unam.mx}}

\pubyear{2014}
\volume{298}  
\pagerange{XX -- YY}
\jname{IAUS\,298 Setting the scene for Gaia and LAMOST}
\editors{S. Feltzing, G. Zhao, N.\,A. Walton \& P.\,A. Whitelock, eds.}

\newcommand{\kmprs}  {\mbox{\rm km\,s$^{-1}$}}

\newcommand{\feh} {\mbox{\rm [Fe/H]}}
\newcommand{\cfe} {\mbox{\rm [C/Fe]}}

\newcommand{\ofe} {\mbox{\rm [O/Fe]}}
\newcommand{\oh} {\mbox{\rm [O/H]}}
\newcommand{\co} {\mbox{\rm [C/O]}}
\newcommand{\nafe} {\mbox{\rm [Na/Fe]}}

\newcommand{\mnfe} {\mbox{\rm [Mn/Fe]}}
\newcommand{\nife} {\mbox{\rm [Ni/Fe]}}
\newcommand{\cufe} {\mbox{\rm [Cu/Fe]}}
\newcommand{\znfe} {\mbox{\rm [Zn/Fe]}}

\newcommand{\alphafe} {\mbox{\rm [$\alpha$/Fe]}}
\newcommand{\teff}  {\mbox{$T_{\rm eff}$}}

\newcommand{\logg}  {\mbox{{\rm log}\,$g$}}

\newcommand{\CI} {C\,{\sc i}}
\newcommand{\OI} {O\,{\sc i}}
\newcommand{\FeI} {Fe\,{\sc i}}
\newcommand{\FeII} {Fe\,{\sc ii}}

\begin{document}

\maketitle

\begin{abstract}
Carbon and oxygen abundances in F and G main-sequence stars
ranging in metallicity from [Fe/H] = $-1.6$ to +0.5 are
determined from a non-LTE analysis of C\,{\sc i} and O\,{\sc i} atomic lines
in high-resolution spectra. Both C and O are good tracers of
stellar populations; distinct trends of [C/Fe] and [O/Fe] as
a function of [Fe/H] are found for high- and low-alpha halo
stars and for thick- and thin-disk stars. These trends and
that of [C/O] provide new information on the nucleosynthesis
sites of carbon  and the time-scale for the chemical enrichment
of the various Galactic components.

\end{abstract}

\firstsection 
\section{Introduction}

Recent $\Lambda$CDM, hydrodynamical simulations of the formation of the Galaxy
(e.g. Zolotov et al. \cite{zolotov10}; Font et al. \cite{font11}; McCarthy et al.
\cite{mccarthy12})
predict the existence of two populations of halo stars. The first one is formed
{\em in situ} in merging and dissipating gas clumps, whereas the other one is
{\em accreted} from 
satellite galaxies. For the upper end of the halo metallicity distribution,
Zolotov et al. (\cite{zolotov10}) find that \alphafe\ 
in the accreted population decreases with increasing [Fe/H] relative to the
near-constant \alphafe\ in the in-situ population. 
This is due to a difference in star formation rate (SFR);
the chemical enrichment proceeds at a slower rate in satellite galaxies, so that
Type Ia SNe start contributing with iron at a lower metallicity.

Evidence for two halo populations\footnote{Selected as stars having a 
total space velocity $V_{\rm total} >  180$\,\kmprs\ relative to the
local standard of rest (LSR).} in the solar neighborhood with different
\alphafe\ trends has been found by Nissen \& Schuster (\cite{nissen10}, hereafter NS10).
The ``high-alpha" stars have $\alphafe \simeq 0.3$, similar to  
thick-disk stars, and extend in metallicity up to $\feh \simeq -0.4$. 
The ``low-alpha" stars show a declining \alphafe -trend from $\sim 0.3$\,dex at
$\feh \simeq -1.6$ to $\sim 0.1$\,dex at $\feh \simeq -0.8$, which is the maximum
\feh\ reached by this population. Furthermore, the Toomre diagram shows that
the high-alpha stars tend to move on prograde orbits, whereas the low-alpha
stars have higher space velocities with respect to the LSR
and an excess of retrograde orbits. These data are consistent with a
scenario where the high-alpha stars have been formed in dissipational
mergers of gas clouds, and the low-alpha stars have been accreted
from satellite galaxies.

The high- and low-alpha halo stars also separate in \nafe , \nife , \cufe ,
and \znfe , but not in \mnfe\ (Nissen \& Schuster \cite{nissen11}).
Furthermore, the high-alpha stars seem to be on average 2 - 3\,Gyr older
than the low-alpha stars (Schuster et al. \cite{schuster12}).

In this paper we report on carbon and oxygen abundances in the two halo populations
and compare with abundances in stars having thin- and thick-disk kinematics.

\section{Carbon abundances}

The abundance of carbon was determined from equivalent widths (EWs) of the 
high-excitation
\CI\ lines at 5052.2 and 5380.3\,\AA\ as measured in high-resolution,
high-S/N VLT/UVES and NOT/FIES spectra. For each star, a 1D model atmosphere
was obtained from the MARCS grid (Gustafsson et al. \cite{gustafsson08}) 
by interpolating to the stellar values of \teff , \logg , \feh , and \alphafe .
The model was used to to derive an LTE carbon abundance. Non-LTE corrections from
Takeda \& Honda (\cite{takeda05}) were afterwards applied, but they are small,
i.e. less than 0.02\,dex.

The atmospheric parameters of the stars were determined spectroscopically
by analysing \FeI\ and \FeII\ lines relative to the same lines in the spectra of
two nearby thick-disk stars, HD\,22879 and HD\,76932, for which \teff\ and \logg\
are well known from colour indices and Hipparcos parallaxes. This enable us to
determine very precise differential parameters and abundances for stars
that belong to the same region of the HR-diagram as the standard stars.
The absolute values may be more uncertain. Thus we have increased the \teff\ scale
by +100\,K relative to the scale used in NS10
to take into account the new IRFM \teff -- colour calibration by
Casagrande et al. (\cite{casagrande10}). This has only a small effect on
\logg , \feh\ (derived from \FeII\ lines), and \alphafe\ (derived from 
neutral lines), but it has a significant effect on C and O abundances
derived from high-excitation atomic lines.

\begin{figure}[t]
\begin{center}
 \includegraphics[width=9cm]{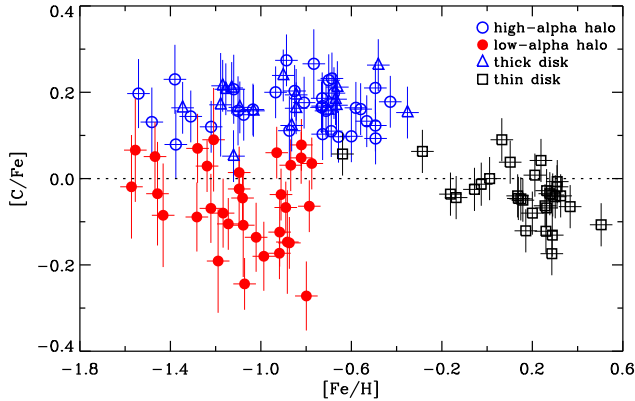} 
 \caption{[C/Fe] versus [Fe/H] with the same classification in high- and
low-alpha halo and thick-disk stars as in NS10
and with thin-disk stars from Nissen (\cite{nissen13}) included.}
   \label{Nissen:fig1} 
\end{center}
\end{figure}

The derived \cfe\ values are shown in Fig. \ref{Nissen:fig1}.
In addition to stars from NS10, we have included  stars with thin-disk kinematics 
from Nissen (\cite{nissen13}), who also used
the $\lambda \lambda 5052, 5380$ \CI\ lines (measured in high-quality
HARPS spectra) to derive C abundances. The error bars shown were
estimated by adding, in quadrature, the errors arising from the EWs 
and those corresponding to the internal 
uncertainties of \teff\ and \logg . At low metallicity and low \alphafe ,
the \CI\ lines are very weak, i.e. $EW \sim$ 2 - 4\,m\AA\ only, so
the error of \cfe\ due to the uncertainty of the EW measurements 
becomes large.
 
\section{Oxygen abundances}

Oxygen abundances were determined from equivalent widths of the $\lambda 7774$
\OI\ triplet lines. The NOT/FIES spectra do not cover this wavelength region
and only a few of the UVES spectra have the \OI\ triplet
included. We have therefore adopted EW measurements from
Ram\'{\i}rez et al. (\cite{ramirez12}), who derived oxygen abundances
for a subset of stars from NS10 using high-resolution spectra obtained
with the 2.7\,m telescope at the McDonald observatory, Keck/HIRES, and 
the MIKE spectrograph at the Magellan Telescope. For the thin-disk stars
we used FEROS spectra obtained with the ESO 2.2\,m telescope. Altogether,
\OI\ triplet data are available for 101 stars out of the 117 stars for which
\CI\ abundances were determined.

In deriving O abundances from the 7774\,\AA\ triplet, non-LTE corrections 
from Fabbian et al. (\cite{fabbian09a}) were applied as described 
by Nissen  (\cite{nissen13}). These corrections are important, even
for differential determinations in
the \teff -range of our sample of stars; the correction of [O/H] 
ranges from about +0.1\,dex at 5300\,K to $-0.1$\,dex at 6300\,K.

\begin{figure}[t]
\begin{center}
 \includegraphics[width=9cm]{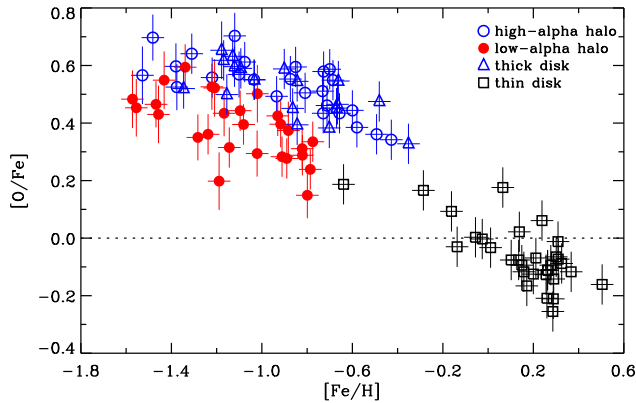} 
 \caption{[O/Fe] versus [Fe/H].} 
   \label{Nissen:fig2}
\end{center}
\end{figure}

The derived \ofe\ values are shown in Fig. \ref{Nissen:fig2}.
The distribution of stars looks much the same as that obtained by
Ram\'{\i}rez et al. (\cite{ramirez12}, Fig. 1, lower panel), except
that they find a flatter distribution of [O/Fe] for the
thick-disk and high-alpha halo stars. This is due to a somewhat stronger dependence
of their non-LTE corrections on [Fe/H] than those of  Fabbian et al. (\cite{fabbian09a}).

\section{Discussion}

\begin{figure}[t]
\begin{center}
 \includegraphics[width=9cm]{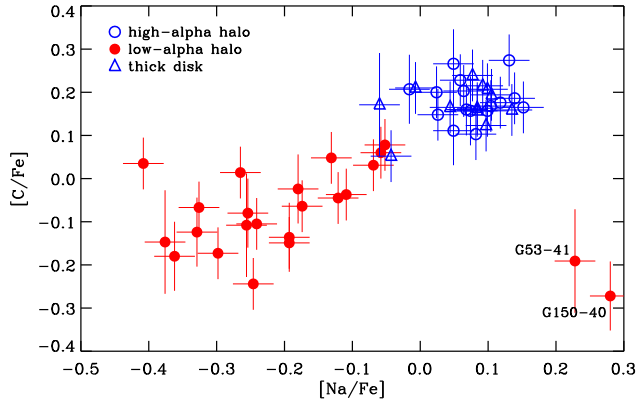}
 \caption{[C/Fe] versus [Na/Fe] for stars with $-1.2 < \feh < -0.7$.} 
   \label{Nissen:fig3}
\end{center}
\end{figure}

As seen from Figs. \ref{Nissen:fig1} and \ref{Nissen:fig2},
the low-alpha halo stars are separated in
\cfe\ and \ofe\ from thick-disk and high-alpha halo stars.
For the latter two populations, the dispersion in \cfe\ and \ofe\
at a given metallicity can be explained in terms of the errors of 
the derived abundances, whereas there seems to be an additional cosmic
dispersion in \cfe\ and \ofe\ for the low-alpha halo population.
In support of a real cosmic scatter, 
Fig. \ref{Nissen:fig3} shows \cfe\ versus \nafe\ (from NS10) for
the metallicity  range $-1.2 < \feh < -0.7$, where the largest
dispersion is present. As seen, there is a striking correlation
between \cfe\ and \nafe\ except for two strongly deviating stars.

As suggested by the $\Lambda$CDM simulations of Zolotov et al. (\cite{zolotov10}),
a possible explanation of the abundance differences seen
in these figures is that the high-alpha stars were born in
the innermost part of the Galaxy  
in a deep gravitational potential with such a high SFR                     
that Type Ia SNe did not contribute significantly with iron 
until a metallicity of $\feh \simeq -0.4$.
Later these stars were dispersed to the halo by merging satellite galaxies.
The low-alpha stars, on the other hand, were formed in
dwarf galaxies with a relatively shallow potential, and hence low SFR, 
so that Type Ia SNe started contributing with iron at a metallicity
around or below $\feh = -1.6$. The reason for the scattter
in \cfe , \ofe , \nafe , and \alphafe\ at a given metallicity could then be that
the various satellite galaxies from which the low-alpha stars were accreted
had different masses and therefore different star formation rates.

The two deviating low-alpha stars in Fig. \ref{Nissen:fig3},
G\,53-41 and G\,150-40, require a special explanation. They  
have high \nafe\ but low \cfe\ and also very low \ofe . Hence, 
they share the so-called Na-O  anti-correlation in second generation
stars in globular clusters, which are thought to be made of gas
polluted by first-generation AGB stars in which the Ne-Na cycle 
has occured. This suggests that some of the halo stars originate
from disrupted globular clusters, as also noted by Ram\'{\i}rez et al. (\cite{ramirez12}).

Figures \ref{Nissen:fig1} and \ref{Nissen:fig2} also give a hint 
for a difference in \cfe\ and \ofe\ between thin- and
thick-disk stars in the overlapping metallicity range
$-0.7 < \feh < -0.2$ although we have only two thin-disk stars
in this range. A systematic difference is well established
in the case of \ofe\ (e.g. Bensby \& Feltzing \cite{bensby06}, Fig. 11b)
and have also been found for the alpha-capture elements, 
Mg, Si, Ca, and Ti (e.g. Adibekyan \cite{adibekyan12}, Fig. 8),
but it was not seen in the case of \cfe\ by
Bensby \& Feltzing (\cite{bensby06}, Fig. 11a), Clearly, we need
more thin-disk stars in the overlapping metallicity range to verify
the possible difference in \cfe\ between the thin- and thick-disk 
populations.

\begin{figure}[t]
\begin{center}
 \includegraphics[width=9cm]{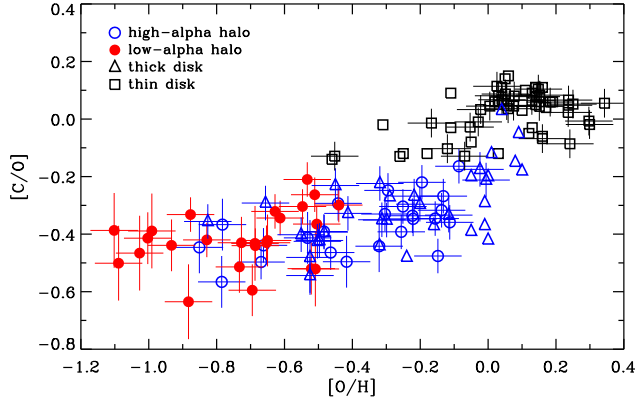}
 \caption{[C/O] versus [O/H]. Thin- and thick-disk stars from 
  Bensby \& Feltzing (\cite{bensby06}) are included without error bars.}
   \label{Nissen:fig4}
\end{center}
\end{figure}

Fig. \ref{Nissen:fig4} shows \co\ versus \oh\ for
the four populations studied in this paper with thin- and
thick-disk stars from Bensby \& Feltzing (\cite{bensby06}) included.
Their C and O abundances are derived from the weak forbidden
$\lambda 8727$ [\CI ] and $\lambda 6300$ [\OI ] lines but agree
very well with our abundances derived from high-excitation
\CI\ and \OI\ lines. 

As seen from Fig. \ref{Nissen:fig4}, there is no systematic
shift in \co\ between high- and low-alpha halo stars. 
This is perhaps surprising, because the time-scale for 
the chemical enrichment of the low-alpha population is
long enough to allow Type Ia SNe to contribute with iron
(our explanation for the low $\alpha$/Fe ratios),
and one could therefore have expected that low- and
intermediate-mass AGB stars had enough time to contribute with
carbon and raise \co\ in the low-alpha stars
to higher values than in high-alpha stars.   
The explanation may be that intermediate-mass (4 - 8 $M_{\rm Sun}$) 
AGB stars contribute very little to $^{12}$C
(Kobayashi et al.  \cite{kobayashi11}), and
that the evolution time-scale of low-mass (1 - 3 $M_{\rm Sun}$) AGB stars
(which do have a high
$^{12}$C yield according to Kobayashi et al.) is longer than the
chemical enrichment time-scale of the low-alpha
population. We conclude that carbon in both high- and low-alpha 
halo stars was made primarely in high-mass stars ($> 10 \,  M_{\rm Sun}$),
but with a metallicity-dependent yield to account 
for the slight increase of \co\ with increasing \oh .

\begin{figure}[b]
\begin{center}
 \includegraphics[width=12cm]{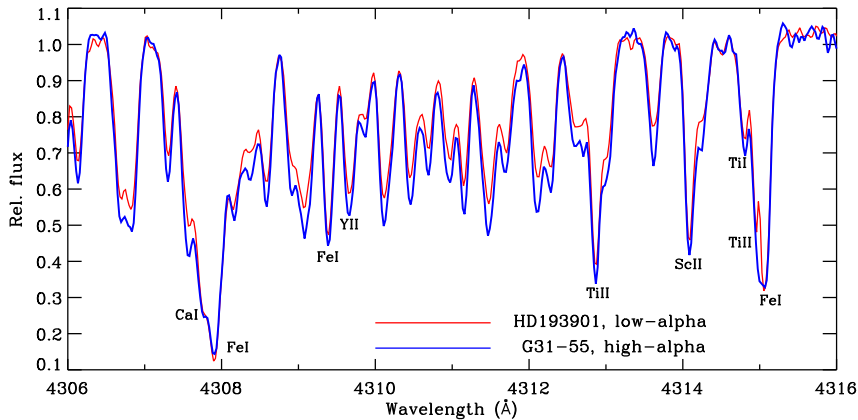}
 \caption{Part of the CH-band in a high- and a low-alpha star with similar
atmospheric parameters. All unmarked lines are due to CH.}
   \label{Nissen:fig5}
\end{center}
\end{figure}

As also seen from Fig. \ref{Nissen:fig4}, thick-disk stars 
have on average the same \co\ as high-alpha halo stars,
but thin-disk stars have systematically higher \co\
in the overlapping metallicity range $-0.5 < \oh < +0.1$.
This offset in  \co\ between thin- and thick-disk stars
was already detected by Bensby \& Feltzing (\cite{bensby06}, Fig. 12),
but is even more clearly seen in our Fig. \ref{Nissen:fig4}.
Probably, the higher \co\ ratios in the Galactic thin disk
are due to low-mass AGB stars, i.e. the
chemical enrichment time-scale of the thin disk 
is long enough to allow low-mass stars to contribute.

As a final remark, we note that the $\lambda \lambda 5052, 5380$ \CI\ lines 
used to derive C abundances in this paper are very weak in metal-poor stars.
Thus, very high S/N, high-resolution spectra are required to derive precise 
C abundances. The somewhat stronger \CI\ lines in the 9000 - 9500\,\AA\ region may
be used instead (Fabbian et al. \cite{fabbian09b}) but they are blended
by telluric lines. As an interesting alternative one may use the CH band. 
Fig. \ref{Nissen:fig5} shows part of this band  
in NOT/FIES spectra of a high-alpha star, G\,31-55 (\teff , \logg , \feh , \alphafe)
= (5738\,K, 4.33, $-$1.10, 0.29) and a low-alpha star HD\,193901 (5756\,K, 4.39, 
$-$1.09, 0.16). As seen, the three \FeI\ lines have similar strengths in the
two stars, whereas all CH lines are significantly weaker in the low-alpha star.
Hence, it may be possible to use the CH band  
in large forthcoming surveys such as HERMES/GALAH and
ESO/4MOST for high precision differential studies of carbon
abundances in stellar populations.

\begin{acknowledgements}
Funding for the Stellar Astrophysics Centre is provided by the
Danish National Research Foundation (Grant agreement no.: DNRF106).
This paper is based on observations made with the Nordic Optical Telescope
on La Palma, and on data from the European Southern Observatory
ESO/ST-ECF Science Archive Facility.
\end{acknowledgements}

\begin{discussion}

\discuss{Nordstr\"{o}m}{Your conclusion that carbon in halo stars and thick-disk stars
was made in high-mass stars - does that also apply to the most metal-poor stars,
[Fe/H] $< -3$?}

\discuss{Nissen}{At very low metallicities, [C/O] increases with decreasing [Fe/H]
(see Fabbian et al. \cite{fabbian09b}). This may be due to a high carbon production
in zero metallicity (Pop. III) stars.}

\discuss{Kobayashi} {The nucleosynthesis yields of \alphafe\ and [C/Fe] do not
depend much on the progenitor metallicity (0.1\,dex for \alphafe ). Your observational
results are consistent with low-mass SNe\,II (13 - 20 $M_{\rm Sun}$), which have lower
\alphafe\, [C/Fe] (and also lower [Cu,Zn/Fe] as you showed in your 2011 paper)
than massive SNeII.}

\discuss{Nissen}{I agree that this is an interesting alternative to our
proposal that the low $\alpha$/Fe ratios in the low-alpha population are caused by
Type Ia SNe starting to contribute with iron at a low metallicity, $\feh < -1.5$.} 

\end{discussion}

\end{document}